\documentclass[preprintnumbers,amsmath,amssymb,pra]{revtex4}
\usepackage[latin1]{inputenc}
\usepackage{dcolumn}
\usepackage{bm}
\usepackage{graphicx}

\def\br{{\bf r}}
\def\bR{{\bf R}}
\def\bk{{\bf k}}
\def\bE{{\bf E}}
\def\m{\mbox{\boldmath $\mu$}}
\newcommand{\ket}[1]{\displaystyle{|#1\rangle}}
\newcommand{\bra}[1]{\displaystyle{\langle #1|}}
\newcommand{\fkjr}{\mathbf{f}(\mathbf{k}j,\mathbf{r})}
\newcommand{\ukj}{1_{\mathbf{k}j}}
\newcommand{\ukjp}{1_{\mathbf{k}'j'}}
\newcommand{\skj}{\sum_{\mathbf{k}j}}
\newcommand{\skkjj}{\sum_{\mathbf{k}\mathbf{k}'jj'}}

\begin{document}

\title{Casimir-Polder forces, boundary conditions and fluctuations}
\author{R. Messina, R. Passante, L. Rizzuto, S. Spagnolo, R. Vasile}
\affiliation{Dipartimento di Scienze Fisiche ed
Astronomiche dell'Universit\`{a} degli Studi di Palermo and CNISM,
Via Archirafi 36, I-90123 Palermo, Italy}

\date{February 22, 2008}

\begin{abstract}
We review different aspects of the atom-atom and atom-wall
Casimir-Polder forces. We first discuss the role of a boundary
condition on the interatomic Casimir-Polder potential between two
ground-state atoms, and give a physically transparent
interpretation of the results in terms of vacuum fluctuations and
image atomic dipoles. We then discuss the known atom-wall
Casimir-Polder force for ground- and excited-state atoms, using a
different method which is also suited for extension to
time-dependent situations. Finally, we consider the fluctuation of
the Casimir-Polder force between a ground-state atom and a
conducting wall, and discuss possible observation of this force
fluctuation.
\end{abstract}

\maketitle

\section{Introduction}
\label{sec:1} Casimir-Polder forces are long-range interactions
between atoms or molecules, or between atoms/molecules and neutral
macroscopic objects, due to their interaction with the
electromagnetic radiation field in the vacuum state
\cite{Milonni94}. In the case of two ground-state atoms the
Casimir-Polder potential energy behaves as $R^{-6}$ for
interatomic distances smaller than typical atomic transition
wavelenghts from the ground state (near zone) and as $R^{-7}$ for
larger distances (far zone) \cite{CT98,CPP95}. In the case of a
ground-state atom and an infinite conducting wall, the potential
energy decreases like $d^{-3}$ in the near zone and $d^{-4}$ in
the far zone, $d$ being the atom-wall distance. The atom-wall
Casimir-Polder force, although very tiny, has been recently
measured with precision, both in the near and in the far zone
\cite{Sukenik1,Sukenik2,DruzhDeKie,HOMC05,OWAPSC07}. The atom-atom
van der Waals/Casimir-Polder energy is still weaker, but
experimental indirect evidences of them exist since a long time,
in agreement with theoretical predictions \cite{VO03}. Direct
measurements of the retarded van der Waals attraction in
mesoscopic systems have been also obtained \cite{CT78,BP99}.
Possible enhancement of such interactions due to the presence of
boundary conditions and resonance effects has been recently
suggested \cite{BD05,Sherkunov07,Tomas07}.

In this paper we review some results, and relative physical
interpretation, about the atom-atom Casimir-Polder force when
boundary conditions are present, and about the atom-wall
Casimir-Polder force. In Section \ref{sec:2} we discuss the force
between two ground-state atoms when a perfectly conducting plate
is present. Section \ref{sec:3} is devoted to the well-known
Casimir-Polder force between a ground- or excited-state atom and a
conducting wall, obtained from a different point of view which can
be also directly extended to dynamical situations. Finally, in
Section \ref{sec:4} we discuss quantum fluctuations of the
atom-wall Casimir-Polder force, their possible measurement and an
estimate of these fluctuations in actual experiments.

\section{Effects of a boundary condition on the interatomic van der Waals/Casimir-Polder force}
\label{sec:2}

In this section we calculate the Casimir-Polder potential between
two neutral ground-state atoms placed in front of a perfectly
conducting infinite plate. For our purposes, it is convenient to
use an effective Hamiltonian obtained from the usual forms of the
interaction Hamiltonian in dipole approximation by a unitary
transformation \cite{PPT98}. For two atoms (A and B) interacting
with the radiation field, this effective Hamiltonian is
\begin{equation}
H_I=-\frac12\sum_{i=A,B}\skkjj
\alpha_i(\omega_k)\mathbf{E}_{\mathbf{k}j}(\br_i)\cdot\mathbf{E}_{\mathbf{k'}j'}(\br_i)
\label{eq:2.1}
\end{equation}
where $\br_i$ is the position of atom $i$, $\alpha_i(\omega_k)$
its dynamical polarizability (assumed isotropic), and
\begin{equation}
\mathbf{E}(\br)=\sum_{\mathbf{k}j}\mathbf{E}_{\mathbf{k}j}(\br)
=i\sum_{\mathbf{k}j}\sqrt{\frac{2\pi
\hbar\omega_k}{V}}\mathbf{f}(\mathbf{k}j,\br)
\left(a_{\mathbf{k}j}-a^{\dagger}_{\mathbf{k}j}\right)
\label{eq:2.2}
\end{equation}
is the transverse displacement field operator,
${\mathbf{f}}(\mathbf{k}j,\br)$ are appropriate mode functions
taking into account the boundary conditions and $j$ is a
polarization index. The effective Hamiltonian \eqref{eq:2.1} is
correct up to the second order in the electron charge $e$. Our
calculation proceeds in two steps. We first obtain the dressed
ground state of one atom (A), and then we evaluate the effective
interaction energy of the second atom (B) with the field
fluctuations generated by the virtual photon cloud dressing the
first atom. The dressed ground state of atom A at the lowest
significant order in the atom-radiation coupling is obtained by
straightforward perturbation theory
\begin{equation}
\hspace{-1.6cm}|\{0_{\mathbf{k}j}\},\downarrow\rangle_D=
|\{0_{\mathbf{k}j}\},\downarrow\rangle-\frac{\pi}{V}\skkjj
\alpha_{A}(\omega_k)\frac{\sqrt{kk'}}{k+k'}\mathbf{f}(\mathbf{k}j,
\br_A)\cdot\mathbf{f}(\mathbf{k'}j',\br_A)|
1_{\mathbf{k}j}1_{\mathbf{k'}j'},\downarrow\rangle \label{eq:2.3}
\end{equation}
where $\mid \downarrow \rangle$ denotes the ground state of atom
A. The interaction energy is obtained from the average value of
the effective interaction Hamiltonian of atom B with the field
(the term with \emph{i}=B of eq. \eqref{eq:2.1}) on the dressed
ground state \eqref{eq:2.3} of atom A, that is
\begin{equation}
\Delta E_{AB}=-\frac{1}{2}\skkjj\alpha_{B}(\omega_k)\
_{D}\langle\{0_{\mathbf{k}j}\},\downarrow|\mathbf{E}_{\mathbf{k}j}(\br_B)
\cdot\mathbf{E}_{\mathbf{k'}j'}(\br_B)
|\{0_{\mathbf{k}j}\},\downarrow\rangle_D \label{eq:2.4}
\end{equation}
After some lengthy algebra, in the far zone we obtain
\begin{equation}\begin{split}
\Delta E_{AB}(R,\bar{R})&=-\frac{23}{4\pi}\hbar c\frac{\alpha_A(0)\alpha_B(0)}{R^7}-\frac{23}{4\pi}\hbar c\frac{\alpha_A(0)\alpha_B(0)}{\bar{R}^7}+\frac{8}{\pi}\hbar c\frac{\alpha_A(0)\alpha_B(0)}{R^3\bar{R}^3(R+\bar{R})^5}\Bigl[R^4\sin^2\vartheta+5R^3\bar{R}\sin^2\vartheta\\
&\,+R^2\bar{R}^2(6+\sin^2\vartheta+\sin^2\bar{\vartheta})+5R\bar{R}^3\sin^2\bar{\vartheta}+\bar{R}^4\sin^2\bar{\vartheta}\Bigr]\label{eq:2.5}\end{split}\end{equation}
where $\alpha_i(0)$ are the static atomic polarizabilities, $R$ is
the distance between the atoms, $\bar{R}$ the distance between one
atom and the image of the other atom reflected on the wall, and
$\vartheta$ ($\bar{\vartheta}$) is the angle between $\bR$
($\bar{\bR}$) and the normal to the plate \cite{PT82,SPR06}.

Equation \eqref{eq:2.5} is a general expression of the
Casimir-Polder potential energy between the two atoms in the
presence of the conducting plate in the far zone. We note in
\eqref{eq:2.5} the presence of three terms: the usual $R^{-7}$
Casimir-Polder potential between the two atoms (as in the absence
of the plate), the $\bar{R}^{-7}$ Casimir-Polder-like interaction
between one atom and the image of the other atom reflected on the
plate, and a term depending from both variables $R$ and $\bar{R}$.
It is possible to show that the interatomic force is attractive
for any spatial configuration of the system. Although expression
\eqref{eq:2.5} has been already obtained by fourth-order
perturbation theory \cite{PT82}, we have now obtained it by a
different and simpler method, which stresses the role of dressed
field fluctuations modified by the presence of the conducting
plate. In fact, we have shown that the Casimir-Polder potential
can be seen as the interaction energy of one atom with the photon
cloud dressing the other atom, which has been modified by the
presence of the conducting plate.

The potential \eqref{eq:2.5} can be also obtained with a different
physical model, based on the properties of the spatial
correlations of vacuum fluctuations which are modified by the
conducting plate \cite{PT93,PPR03}. In fact, the zero-point
electromagnetic field induces a fluctuating dipole moment on each
atom which is related to the fluctuating zero-point field by
(assuming an isotropic atom)
\begin{equation}
\mu_\ell (\bk j)=\alpha(\omega_k)E_\ell(\bk j,\br) \label{eq:2.6}
\end{equation}
Because vacuum fluctuations are spatially correlated, the induced
dipole moments are correlated too. We now let the two induced
dipoles interact through the classical potential between
oscillating dipoles \cite{SPR06}
\begin{eqnarray}
\hspace{-1.2cm}V_{lm}(k,R,\bar{R})&=& V_{lm}(k,R)-\sigma_{lp}V_{pm}(k,\bar{R})
\nonumber \\
&=&(\nabla^{2}\delta_{lm}-\nabla_{l}\nabla_{m})^{R}
\frac{\cos\bigl(kR\bigr)}{R}-\sigma_{lp}(\nabla^{2}\delta_{pm}-\nabla_{p}
\nabla_{m})^{\bar{R}}\frac{\cos\bigl(k\bar{R}\bigr)}{\bar{R}}
\label{eq:2.7}
\end{eqnarray}
where the first term is the interaction between the two induced
dipole moments as in the absence of the plate, and the second term
takes into account the presence of the conducting plate with a
contribution due to the interaction between one atom and the image
of the other atom. In \eqref{eq:2.7}, which is indeed related to
the real part of the field Green's function as obtained in
\cite{WS84} for our planar geometry, $\sigma_{lp}$ is the
reflection matrix with respect to the plate (for a non-perfectly
conducting plate, it would also depend on the wavevector), and the
superscripts $R,\bar{R}$ indicate the variables with respect to
which derivatives are taken. The interaction energy between the
two atoms is thus
\begin{eqnarray}
V_{AB}&=&\skj \bra{0}\mu_\ell^A (\bk j)\mu_m^B (\bk j) \ket{0}
V_{\ell m}(k,R,\bar{R}) \nonumber \\
&=&\skj\alpha_A(\omega_k)\alpha_B(\omega_k)\bra{0}E_\ell(\bk
j;\br_A)E_m(\bk j;\br_B)\ket{0}V_{\ell m}(k,R,\bar{R})
\label{eq:2.8}
\end{eqnarray}
where the spatial correlation function of the electric field
operator appears, which, similarly to $V_{lm}$, depends on the
boundary conditions. After straightforward algebra, evaluation of
\eqref{eq:2.8} again yields eq. \eqref{eq:2.5}. This gives a second
physically transparent interpretation of the origin of the
Casimir-Polder energy between the two atoms with boundary
conditions present. It also gives support to the phenomenological
model used and indicates that it could be applied also to more
complicated boundary conditions. The effect of boundary conditions
on the Casimir-Polder interaction between atoms, when approximated
as harmonic oscillators, has been also considered in the
literature at zero and finite temperature using Green's functions
techniques \cite{MN73,BLM01}. An interpretation of the van der
Waals potential in terms of dipole images has been also given in
\cite{McLachlan64}.

\section{The atom-wall Casimir-Polder force for ground- or excited-state
atoms} \label{sec:3}

We now consider the Casimir-Polder force between one atom and a
neutral perfectly conducting infinite plate. The atom-plate
Casimir-Polder interaction is well-know in the literature, both
for ground- and excited-state atoms \cite{Haroche91,Barton87}; we
shall however derive it from a different point of view. In the
multipolar coupling scheme, the atom-field interaction is given by
$H_I = - \m \cdot \bE (\br )$, where $\m$ and $\bE(\br )$ are
respectively the atomic dipole moment and the transverse
displacement field, $\br$ being the position of the atom (we do
not use the effective Hamiltonian \eqref{eq:2.1} because it is
limited to systems in their ground state). We consider a two-level
atom with transition frequency $ck_0$ between the two levels $1$
and $2$, interacting with the field by the multipolar coupling
Hamiltonian. Due to the atom-field interaction, the unperturbed
states become dressed; in our approach, the atom-plate
Casimir-Polder interaction is obtained from the average value of
$H_I$ on the first-order dressed state $\mid
\{0_{\mathbf{k}j}\}\downarrow (\uparrow ) \rangle_D$ as
\cite{CPP83}
\begin{equation}
\Delta E_{g(e)}=\frac 12 \
_{D}\langle\{0_{\mathbf{k}j}\},\downarrow \left( \uparrow
\right)\mid H_I \mid \{0_{\mathbf{k}j}\},\downarrow \left(
\uparrow \right)\rangle_D \label{eq:3.1}
\end{equation}
where $\downarrow$ and $\uparrow$ refer respectively to a
ground-state and an excited-state atom. The interaction energy is
then
\begin{equation}
\Delta E_{g(e)} = -\frac {4\pi}V \sum_{\bk j} \frac k{k\pm k_0}
\left( \m_{21} \cdot \fkjr \right)^2 \label{eq:3.2}
\end{equation}
where $\fkjr$ are the field mode functions evaluated at the
position of the atom; the plus sign refers to a ground-state atom
and the minus sign to an excited-state atom  (the known
difficulties of the two-level model in dealing with frequency
shifts near an interface \cite{Barton74} do not seem relevant in
our case; our results, derived in the multipolar coupling scheme,
can be straightforwardly generalized to the case of a multilevel
atom, obtaining an expression analogous to \eqref{eq:3.2} with a
summation over all atomic states). For isotropic atoms, after
summation over the orientations of the atomic dipole, the
expressions for the Casimir-Polder force for a ground- and an
excited-state atom are
\begin{eqnarray}
F_g &=& -\frac {\mu^2}{12\pi d^4} \left[ 8k_0 d -6\left(2k_0^2d^2-1\right) f(2k_0d)-4k_0d\left(2k_0^2d^2-3\right) g(2k_0d) \right]\label{eq:3.3}\\
F_e &=& \frac {\mu^2}{12\pi d^4} \left[ 8k_0 d -6\left(2k_0^2d^2-1\right) \left( f(2k_0d)-\pi \cos (2k_0d) \right)-4k_0d\left(2k_0^2d^2-3\right) \left( g(2k_0d) -\pi \sin(2k_0d) \right) \right] \label{eq:3.4}\end{eqnarray}
where $f(z)$ and $g(z)$ are the auxiliary functions of the sine-
and cosine-integral functions \cite{AS65}. The energy shifts in
\eqref{eq:3.2} can be also expressed in terms of the field Green's
functions, which include the summation over the field modes
\cite{WS84,WS85}; also, the contributions from vacuum fluctuations
and radiation reaction (which includes the dipole image field) can
be explicitly separated \cite{MJH90}. For a ground-state atom, Eq.
\eqref{eq:3.3} gives an attractive force for any atom-wall
distance. For an excited-state atom, Eq. \eqref{eq:3.4} shows
spatial oscillations of the force with a periodicity related to
$k_0$, yielding spatial regions where the force is attractive and
regions where the force is repulsive. Although these results are
already known in the literature (see e.g. \cite{Haroche91,HS91}),
our approach gives a physical picture of the atom-wall interaction
as the average interaction energy of the atom with the field
fluctuations associated with its dressing photon cloud, whose
properties depend on the presence of the wall. Also, this approach
allows a conceptually easy evaluation of dynamical, i.e.
time-dependent, atom-plate Casimir-Polder forces in nonequilibrium
situations such as during the dynamical self-dressing of the atom
\cite{VP08}. A similar approach has been already used for the
dynamical Casimir-Polder force between two atoms \cite{RPP04};
time-dependence of the force for an atom in a superposition of
energy states has been also recently considered \cite{BKWD04}.

\section{Fluctuations of the atom-wall Casimir-Polder force}
\label{sec:4}

The atom-wall Casimir-Polder force in Section \ref{sec:3} is an
average value of the force, and quantum fluctuations are expected.
Atom-wall force fluctuations have been already considered (in the
far zone only) by a Langevin-like equation for the position of the
atom \cite{WKF02}. In this Section we evaluate the atom-wall
Casimir-Polder force fluctuation for a ground-state atom by using
an operator directly associated to the force acting on the atom;
our method can be used both in the near and in the far zone. The
atom-radiation coupling is described using the effective
Hamiltonian \eqref{eq:2.1}. The first step is to define a quantum
operator directly associated to the force experienced by the atom,
in order to easily calculate its fluctuation. This operator is
defined by $F=-\partial H_I/\partial d$, i.e. by taking minus the
formal derivative of the operator $H_I$ with respect to the
atom-wall distance $d$, inspired by a classical quasi-static
analogy (we neglect the atomic translational degrees of freedom).
The mean value of this operator on the ground state of our system
gives back the correct value of the force of the previous Section
\begin{equation}
F(d)=-\frac{3\hbar c\alpha}{2\pi d^5}
\label{eq:4.1}
\end{equation}
(far zone) where $d$ is the atom-wall distance and $\alpha =
2\mu_{21}^2/3\hbar \omega_0$ the static polarizability of the
atom. Once we have a force operator, we can square it in order to
obtain the force fluctuation $\Delta F = ( \langle F^2 \rangle
-\langle F \rangle^2 )^{1/2}$, where average values are taken on
the ground state of the system. Unfortunately, a direct evaluation
of $\Delta F$ yields a result containing non-regularizable
ultraviolet divergences. Similar divergences were obtained for the
fluctuation of the macroscopic Casimir force between two parallel
infinite conducting plates. In order to solve this problem we have
followed the approach initially proposed by Barton
\cite{Barton1,Barton2,Barton94}, by taking into account the fact
that any force measurement has a finite duration and thus
introducing a time-averaged force operator. This operator is an
average of the force operator in the Heisenberg representation
over a weight function $f(t)$ describing the instrumental
response, that is
\begin{equation}
\overline{F}=\int_{-\infty}^{+\infty}dt\,f(t)
e^{\frac{i}{\hbar}Ht}Fe^{-\frac{i}{\hbar}Ht}
\label{eq:4.2}
\end{equation}
where $F$ is the force operator defined above and $H$ the
Hamiltonian of the system. This time-averaging procedure
introduces a cut-off in the frequency integration related to the
duration of the measure, which makes the force fluctuation finite.
Using this technique and choosing a Lorentzian of width $T$ as the
instrumental response function
$f(t)=\frac{1}{\pi}\frac{T}{t^2+T^2}$, we obtain
\begin{equation}
\Delta\overline{F}^2=\frac{1}{2}\skkjj\Bigl|\bra{0}F\ket{\ukj\ukjp}
\Bigr|^2\bigl|g(\omega_k+\omega_{k'})\bigr|^2 \label{eq:4.3}
\end{equation}
where $g(\omega)=\exp(-\omega T)$ is the Fourier transform of
$f(t)$. It is thus evident that the use of the time-averaged
operator $\overline{F}$ naturally introduces a frequency cutoff,
making the force fluctuation finite. This is physically reasonable
since an instrument characterized by a response time $T$ is not
able to resolve processes with frequencies higher than $T^{-1}$.
By considering the relative fluctuation of the force (ratio
between $\Delta\overline{F}$ and the mean force), we find two
different regimes according to the ratio between the atom-wall
distance $d$ and $cT$. When $d<<cT$ (long measurements), we get
\begin{equation}
\frac{\Delta\overline{F}}{|\bra{0}F\ket{0}|} \sim
\left(\frac{d}{cT}\right)^6 \label{4.5}
\end{equation}
while when $d>>cT$ (short measurements), we obtain
\begin{equation}
\frac{\Delta\overline{F}}{|\bra{0}F\ket{0}|} \sim
\left(\frac{d}{cT}\right)^5 \label{4.6}
\end{equation}
In the former case, force fluctuations are hardly observable; in
the latter case, the force largely fluctuates around its mean
value. In order to analyse the observability of the fluctuations,
we can consider typical parameters from actual experimental setups
for the atom-wall force measurements as in
\cite{Sukenik1,Sukenik2}. In this case we have $d \sim 1 \,\mu
\mbox{m}$; hence the duration $T$ distinguishing the two regimes
is $\sim 10^{-14}\,\mbox{s}$ \cite{MP07}. Theqrefore, observation
of the quantum fluctuations of the Casimir-Polder force requires
very short measurements compared to measurement times in recent
experiments (for example, a rough estimate in the experiments in
References \cite{Sukenik1,Sukenik2} gives $T \sim 10^{-5} \,
\mbox{s}$, much longer than the time required to observe the force
fluctuation). Observation of the fluctuations of the
Casimir-Polder force requires much shorter measurement times, or
to envisage situations where the force and its fluctuations could
be significantly enhanced (for example through resonance effects
for atoms in excited states).

\section{Conclusions}
\label{sec:5}

We have reviewed several different aspects of Casimir-Polder
forces. In Section \ref{sec:2} we have considered how boundary
conditions can affect the Casimir-Polder interaction between two
ground-state atoms, and given a transparent physical
interpretation of the results in terms of zero-point fluctuations
and image dipoles. In Section \ref{sec:3} we have considered the
Casimir-Polder force between an atom and a conducting wall,
obtained as the interaction energy of the atom with its dressing
field fluctuations. Finally, in Section \ref{sec:4} we have
analysed quantum fluctuations of the Casimir-Polder force between
a ground-state atom and a conducting wall, and discussed their
observability.

\section{Acknowledgments}
\label{sec:6} The authors wish to thank G. Compagno and F. Persico
for stimulating discussions on the subject of this paper. Partial
support by Ministero dell'Universit\`{a} e della Ricerca
Scientifica e Tecnologica and by Comitato Regionale di Ricerche
Nucleari e di Struttura della Materia is also acknowledged.

\end{document}